# Mining individual daily commuting patterns of dockless bike-sharing users: a two-layer framework integrating spatiotemporal flow clustering and rule-based decision trees

Caigang Zhuang[a,b], Shaoying Li[b], Haoming Zhuang[a,]*  Xiaoping Liu[a,]*

[a] School of Geography and Planning, Sun Yat-sen University, Guangzhou, China

[b] School of Geography and Remote Sensing, Guangzhou University, Guangzhou, China

## Abstract

The rise of dockless bike-sharing systems has led to increased interest in using bike-sharing data for sustainable transportation and travel behavior research. However, these studies have rarely focused on the individual daily mobility patterns, hindering their alignment with the increasingly refined needs of active transportation planning. To bridge this gap, this paper presents a two-layer framework, integrating improved flow clustering methods and multiple rule-based decision trees, to mine individual cyclists' daily home-work commuting patterns from dockless bike-sharing trip data with user IDs. The effectiveness and applicability of the framework is demonstrated by over 200 million bike-sharing trip records in Shenzhen. Based on the mining results, we obtain two categories of bike-sharing commuters (74.38% of *Only-biking commuters* and 25.62% of *Biking-with-transit commuters*) and some interesting findings about their daily commuting patterns. For instance, lots of bike-sharing commuters live near urban villages and old communities with lower costs of living, especially in the central city. *Only-biking commuters* have a higher proportion of overtime than *Biking-with-transit commuters*, and the Longhua Industrial Park, a manufacturing-oriented area, has the longest average working hours (over 10 hours per day). Moreover, massive users utilize bike-sharing for commuting to work more frequently than for returning home, which is intricately related to the over-demand for bikes around workplaces during commuting peak. In sum, this framework offers a cost-effective way to understand the nuanced non-motorized mobility patterns and low-carbon trip chains of residents. It also offers novel insights for improving the bike-sharing services and planning of active transportation modes.

**Keywords**: Dockless Bike-sharing; Spatiotemporal Flow Clustering; Rule-based Decision Trees; Commuting Pattern; Data Mining; Sustainable Transportation

1. Introduction

Compared with other modes of mobility, cycling is considered an eco-friendly, healthy, and sustainable mode of transportation, which has a beneficial effect on reducing traffic congestion, energy consumption, and air pollution (DeMaio, 2009; Handy et al., 2014). In the past decade, the spread of the bike-sharing programs has further expanded the benefits of cycling. For example, the convenience of mobile payments and the flexibility of station-less rental services have made dockless bike-sharing, one of the innovative bike-sharing systems, widely accepted and utilized worldwide (Heinen et al., 2010; Zhang & Mi, 2018; Si et al., 2019). These bike-sharing programs have empowered cycling to play an essential role in solving the first-and-last-mile trip problem and enhancing the resilience of urban transportation networks (Fishman, 2016; Teixeira et al., 2021; L. Cheng et al., 2022). They are also regarded as a means of providing cleaner transportation in the building of smart cities (Eren & Uz, 2020). Therefore, how to increase the cycling willingness of residents to promote the development of non-motorized and active transportation has received extensive research attention.

In the early years, relevant studies were conducted based on travel survey data which have the limitations of high cost, low timeliness, and small sample size (S. Li et al., 2021). With the advent of big data era and new bike-sharing systems, the availability of GPS datasets from bike-sharing operators have opened opportunities for cycling-related research. Existing literature has proven that such GPS trajectory data have the advantages of objectivity, high spatiotemporal resolution, and large sample volume (Lu & Liu, 2012). Meanwhile, many scholars have used these data for cycling influence mechanisms analysis (Shen et al., 2018; Ma et al., 2020; F. Gao et al., 2021; D. Wang et al., 2024; Zhu et al., 2024), travel pattern mining (X. Zhou, 2015; Du et al., 2019; Cao et al., 2020; F. Gao et al., 2022; X. Xu et al., 2023; Dzięcielski et al., 2024), trip purpose inference (Xing et al., 2020; S. Li et al., 2021; Ross-Perez et al., 2022), socio-economic benefit assessment (Zhang & Mi, 2018; Luo et al., 2019; Y. Wang & Sun, 2022; Lv et al., 2024), and transportation equity evaluation (Meng & Brown, 2021; Z. Zhou & Schwanen, 2024; B. Wang et al., 2024). For instance, Shen et al. (2018) explored the factors influencing bike usage based on nine consecutive days of bike-sharing trip records in Singapore, and found that high land use mixtures, easy access to public transportation, and more available cycling facilities are positively correlated with bike-sharing usage. In a study that used a week of bike-sharing data collected in Shenzhen, Li et al. (2021) proposed a framework for inferring the trip purpose of cyclists based on gravity models and Bayesian rules, and revealed the spatiotemporal patterns of nine categories of travel activities. Additionally, Y. Zhang & Mi (2018) extracted bike-sharing usage frequency and trip distances in Shanghai, and estimated the environmental benefits of bike-sharing on emission reduction. In a recent study, B. Wang et al. (2024) used the Gini coefficient and panel regression method to reveal serious inequities in bike-sharing usage within disadvantaged communities (with low education levels and high crime rates) in Chicago, although built environment factors such as park space were found to have a positive effect on promoting bike-sharing trips in these communities. These studies are meaningful as they deepen our understanding of the role that bike-sharing plays in urban transportation and residents' travel behaviors.

However, the aforementioned research based on bike-sharing trip data has rarely focused on the daily travel habits of individual cyclists, despite some leveraging datasets that contain user IDs. To date, the most relevant research has been conducted by a limited number of scholars who attempt to explore the travel characteristics of different user groups, utilizing user attributes information (e.g., age, gender, or membership) available within the docked bike-sharing trip dataset (X. Zhou, 2015; Y. Yao et al., 2019; Pellicer-Chenoll et al., 2021; Reilly et al., 2022). For example, Zhou (2015) constructed bike flow similarity graphs and used community detection techniques to discover the different travel trends for customers and subscribers in Chicago. Pellicer-Chenoll et al. (2021) explored the docked bike-sharing usage patterns of male and female users on weekdays in Valencia using graph theory and Voronoi spaces, and they found that female users are more concerned with factors related to travel safety. Although these studies contributed to the insights into the differences in travel patterns within the cycling groups, these methods are not applicable to most bike-sharing trip dataset that include few individual attributes for privacy concerns. Moreover, the relevant studies mentioned above merely categorizes cycling groups based on the user attributes, rather than extracting daily bicycle mobility patterns at the individual level.

Notably, mining the individual daily mobility patterns of bike-sharing users at a finer resolution holds significant implications for the increasingly focus on sustainable and active transportation planning (Ferretto et al., 2021). For instance, it can serve as a low-cost, high-coverage technique to complement traditional, expensive, and less comprehensive travel surveys, assisting urban planners and bike-sharing operators in understanding residents' daily low-carbon trip chains and cycling needs. Furthermore, if bike-sharing users' residential and workplace information can be identified from individual daily mobility patterns, it would enable the integration of various socioeconomic data (e.g., housing price) to explore fine-scale studies of cycling behaviors considering population differentiations (Y. Xu et al., 2018; Wu et al., 2023), thereby providing decision-making basic for the building of human-oriented and bicycle-friendly environments.

So far, there have been some studies proposing methodological frameworks for mining individual daily mobility patterns based on specific geotagged big data, such as cellphone call detail records (CDR) data (Kung et al., 2014; Jiang et al., 2017; Yin et al., 2021), check-in data (Z. Cheng et al., 2011; L. Li et al., 2013; Niu & Silva, 2023; Wu et al., 2023), and smart card data (Sari Aslam et al., 2019; Zhang et al., 2020; Huang et al., 2024). However, these studies' utilized geotagged data do not include any fields related to cycling trips, thus we cannot identify individual cycling patterns from their mining results. Additionally, due to differences in data features, travel characteristics, and influencing factors, dockless bike-sharing trip data are not suitable as inputs for these frameworks. For example, Jiang et al.

(2017) developed an integrated pipeline that can parse, filter, and expand the CDR data to extract human mobility patterns. However, since most bike-sharing trip data only record cycling origins and destinations (ODs), rather than capturing continuous trajectory like CDR data, the relevant extraction methods are not suitable for bike-sharing data. Moreover, in a study leveraging Twitter check-in data, Z. Cheng et al. (2011) proposed a recursive grid search method to detect users' homes and subsequently analyze their mobility patterns. Although it is feasible to reconstruct bike-sharing data into check-in-like data by delineating the origin and destination of each trip, this approach leads to the loss of key cycling attribute (e.g., trip distance and duration). In other words, using bike-sharing data to check-in-based mining methods can only exploit partial data information. Compared with the geotagged data mentioned above, the features of smart card data are closer to those of bike-sharing data. Several studies have proposed methods based on such data to extract users' daily activities, such as a heuristic model developed by Sari Aslam et al. (2019) for detecting the residence and workplace of individuals, decision tree methods presented by Y. Zhang et al. (2020) and Huang et al. (2024) for identifying the stay areas and daily activity spaces of individuals. Nevertheless, noted that the locations of transit stations in the smart card data are fixed, which is significantly different from dockless bike-sharing. Moreover, the travel characteristics and influencing factors of public transportation also differ from those of cycling (e.g., shorter trip distances, more affected by weather and etc.). Hence, there are also limitations in using the such extraction methods for dockless bike-sharing data.

In summary, existing studies on bike-sharing have rarely focused on mining mobility patterns at the individual level, and the methods for individual mobility pattern extraction based on other geotagged data are not suitable for dockless bike-sharing data. To address these gaps, this paper will present a two-layer framework that aims to capture the most dominant daily mobility pattern (i.e., home-work-commuting) of individual dockless bike-sharing users. Specifically, in Layer 1, we develop flow clustering methods that improved spatiotemporal constraints tailored to the travel characteristics of bike-sharing. This enhancement allows us to derive spatiotemporal flow clusters that effectively representing individual daily travel trajectories, from the biking records that lack accurate geocoding information. However, these trajectories identified in Layer 1 still lack activity semantics. Therefore, in Layer 2, we further establish rule-based decision trees that incorporate round-trip journeys, working hours, and public transport transfers for identifying daily commuting trips from individual spatiotemporal flow clusters. Based on individual daily commuting behaviors, we divide bike-sharing users into *Only-biking* and *Biking-with-transit* commuters. To examine the effectiveness and applicability of this two-layer framework, this paper conducts an empirical study using comparative analysis and residence location testing in Shenzhen, China, a metropolis with over one million daily bike-sharing trips. Finally, based on the mining results of our framework, we analyze the daily commuting characteristics and spatiotemporal patterns of individual bike-sharing users, and discuss some meaningful findings and policy implications.

## 2. Study area and dataset

*2.1 Study area*

Shenzhen is located in the Guangdong-Hong Kong-Macao Bay Area, which is one of the most densely populated and economically prosperous regions in China. By the end of 2021, Shenzhen has a permanent population of over 17 million and a regional gross domestic product (GDP) of over 390 billion EUR or 470 billion USB (based on the average exchange rate of 2021) (Guangdong Statistics Bureau, 2021). The high-frequency population mobility and booming socioeconomic activities are accompanied by huge travel demand. To promote low-carbon transportation and encourage more residents to adopt eco-friendly mobility, Shenzhen has made considerable efforts to expand its public transportation system (11 metro lines and 927 bus lines have been opened as of 2021, Transportation Bureau of Shenzhen, 2021) and non-motorized transport infrastructure (e.g., the introduction of bike-sharing systems and the planning of dedicated bike lanes). According to the latest official statistics, the proportion of non-motorized trips (i.e., cycling and walking) in Shenzhen has consistently remained above 50%, while the share of green transportation modes (including public transit and non-motorized travel) has reached 77.42% in recent years (Shenzhen Government Online, 2021).

Among them, the dockless bike-sharing system was first introduced to Shenzhen in 2016. After the initial period of market dominance and the subsequent period of policy regulation, bike-sharing services have been integrated into the daily mobility of local residents. As of July 2022, Shenzhen has over 41,000 dockless bike-sharing with an average of approximately 1.29 million daily trips (Statistics Bureau of Shenzhen, 2022). Usage hotspots are mainly located in financial centers, around high-density residential areas and within industrial parks in districts such as Futian, Nanshan, Luohu, Bao'an, and Longhua (Fig.1). These hotspot areas are typically characterized by convenient connections to public transit, well-developed non-motorized transport infrastructure, or close proximity between residential and employment areas (F. Gao et al., 2022). The substantial volume of bike-sharing trips offers a rich resource for this paper to mine individual daily cycling commuting patterns.

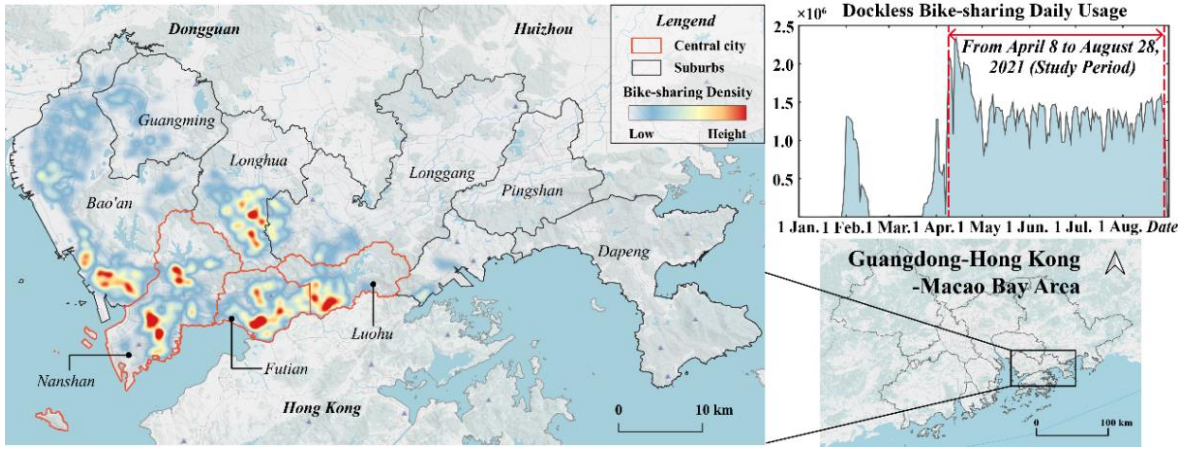

**Fig. 1** Spatial and temporal distribution of raw dockless bike-sharing data in the study area.

*2.2 Data description*

The dockless bike-sharing dataset used in this study is collected from the Shenzhen government data open platform (https://opendata.sz.gov.cn/). The dataset stores over 244 million riding records between January and August 2021, which includes the user IDs and the coordinates and time information of OD. Notably, all user IDs are encrypted and no personal privacy information can be obtained (Table 1). In addition, considering the integrity and continuity of the raw dataset, we finally extract approximately 146 million records that occurred on all weekdays between April 8 and August 28, 2021 for the empirical study below (Fig.1). The exclusion of records during weekends and holidays is due to the substantial occurrence of non-commuting trips during these periods, which could increase data noise.

Moreover, this study also acquire historical daily weather data for the study period (https://lishi.tianqi.com/shenzhen/), public transportation station data (including location and passing bus or metro routes information) in 2021 (https://lbs.amap.com/), and the latest residential land use data (obtained from Shenzhen Municipal Housing and Construction

Bureau, https://zjj.sz.gov.cn/fwzljgcx). The weather data is applied to extract active bike-sharing users, while the station data is employed to identify the transfer behaviors in individual daily commuting trips. As for residential land parcels data, it is used to validate the identification results of individual home locations of bike-sharing users. The details will be presented in Section 3.

Table 1 Example of dataset.

| User ID | Starting Time | Origin | Ending Time | Destination |
|---|---|---|---|---|
| 9fb2d1ec6142ace4d74 05b******** | 2021/01/30 13:19:32 | 114.0082,22 .6392 | 2021/01/30 13:23:18 | 114.0104,22 .6348 |
| 1184eecf9f54441b389 bcf******** | 2021/01/31 23:49:12 | 113.8540,22 .5884 | 2021/01/31 23:54:37 | 113.8528,22 .5840 |
| 30a457b24805ffab03b 9c4******** | 2021/01/30 13:09:10 | 114.0228,22 .6506 | 2021/01/30 13:23:24 | 114.0406,22 .6404 |

## 3. Method

The flowchart of this paper depicted in Fig.2. Initially, "Data filtering and active bike-sharing users identification" aims to eliminate abnormal cycling records and bike-sharing inactive users to enhance data quality. The novel two-layer framework is then employed to mine individual daily cycling commuting patterns, which is the core of this paper. Afterwards, "Framework evaluation and validation" intends to examine the effectiveness and applicability of our framework through the comparative analyses of flow clustering methods and the

10/58

testing of users' residences. Finally, we aggregate and visualize the mining results of individual bike-sharing users to reveal their commuting regularities and spatiotemporal patterns.

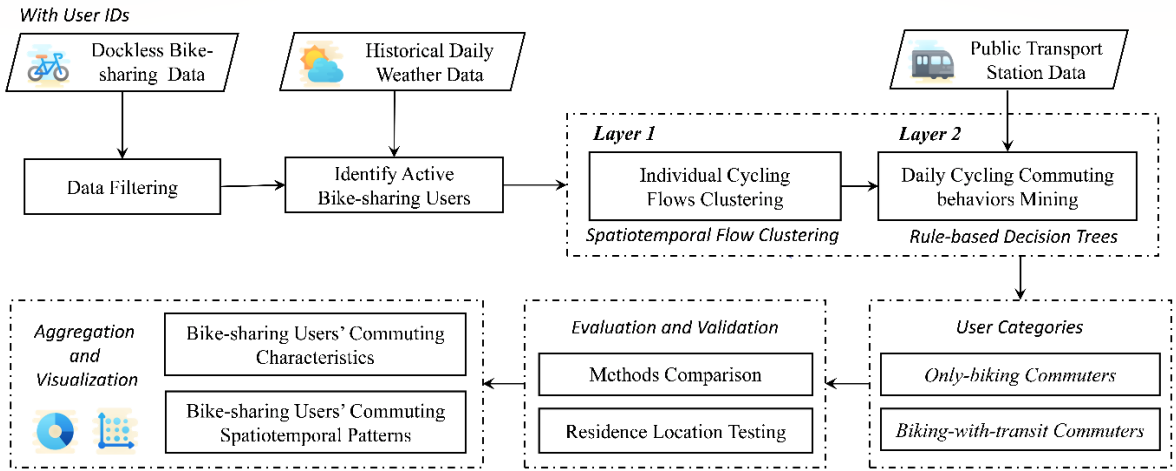

**Fig.2** Research flowchart of this paper.

*3.1 Data filtering and active bike-sharing users' identification*

To ensure the accuracy and authenticity of the bike-sharing data applied to this study, anomalous or redundant records need to be cleaned. First, with reference to existing studies (F. Gao et al., 2021; Shen et al., 2018), the unrealistic long-or-short distances or durations cycling trips due to GPS drifting or misoperations are eliminated. Afterwards, we aggregate the trips of each bike-sharing user based on user IDs and exclude duplicate records existing within the same user. Ultimately, about 2.53 million users' cycling records are extracted.

Moreover, by tallying the number of active days (i.e., have at least one trip record within a day) for users in the filtered weekdays data (Fig.3(a)), we observe the issues of data sparsity. While some users rely heavily on bike-sharing for their daily activities, while others, such as tourists or occasional cyclists, contribute sporadically to the dataset. For the latter, their limited trips cannot adequately capture their daily cycling habits. Hence, it is necessary to exclude these sparse users to ensure a meaningful dataset for revealing relatively complete mobility patterns of individuals.

In a related study, Xu et al. (2018) defined active users in CDR data as those with at least one record for at least half of the study period. However, for bike-sharing dataset, it is crucial to consider the influence of weather on daily cycling. Existing studies have indicated that rainfall can significantly restrict cycling during commuting hours, as people tend to choose other safer transport modes (Reiss & Bogenberger, 2016; Shen et al., 2018). Similarly, in the dataset we used, bike-sharing usage is observed to be lower on drizzly and rainy days (Fig.3(b)). Hence, expanding on the approach of Xu et al. (2018), we exclude the rain-impacted weekdays to establish the threshold for active bike-sharing users, calculated as half of the total weekdays during the data collection period minus the number of drizzly and rainy days. In this study, with 100 weekdays and 21 drizzly and rainy days, the threshold is set at 29 days, thus identifying approximately 0.75 million active bike-sharing users for subsequent processing.

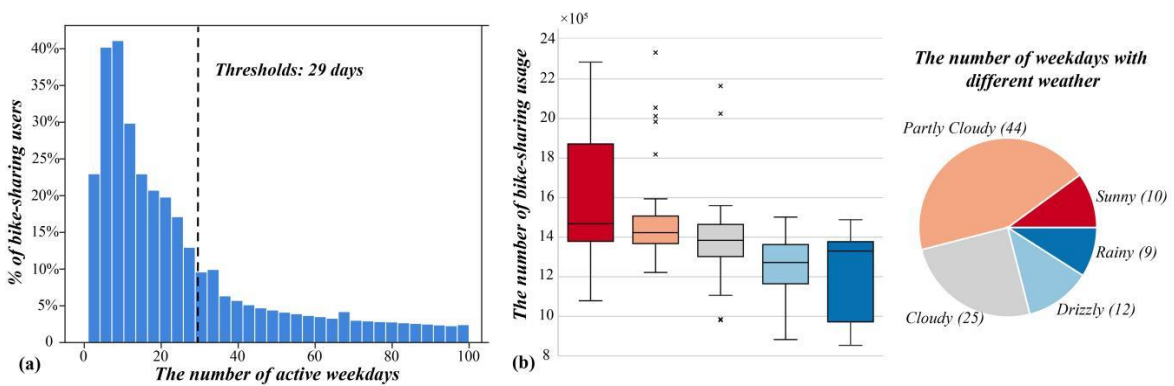

**Fig.3** (a) Histogram of the number of active days for bike-sharing users on weekdays; (b) Relationship between weekdays bike-sharing usage and different weather.

## 3.2 Two-layer framework for individual daily commuting patterns of bike-sharing users

Fig. 4 shows the diagram of two-layer framework. In Layer 1, to address the lack of geocoding information in the records of dockless bike-sharing, we propose an improved spatiotemporal flow clustering method based on the travel characteristics of bike-sharing[*]. This method can extract individual flow clusters representing the daily cycling trajectories of each user. In Layer 2, multiple rule-based decision trees that integrate round-trip journeys, working hours, and public transport transfers is built to identify of bike-sharing commuting behaviors from the individual flow clusters extracted in Layer 1. The identification results are used to classify different categories of bike-sharing commuters.

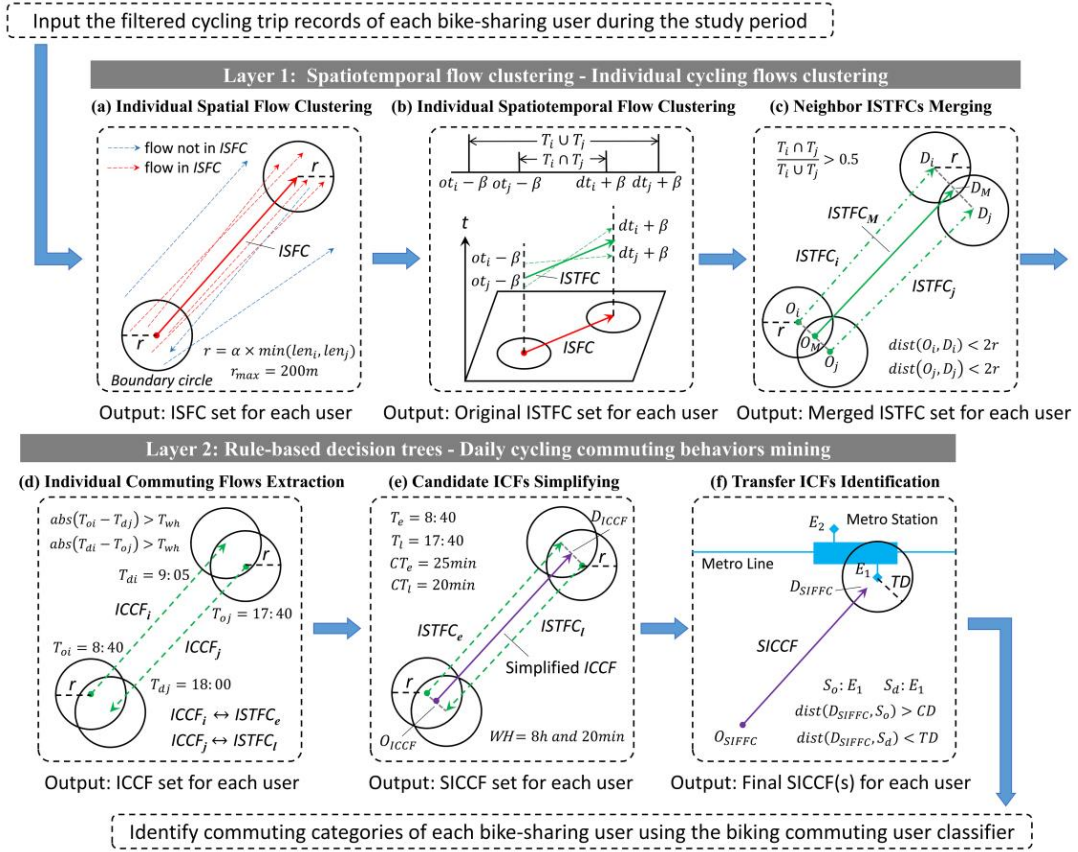

---

[*] The sample data and code for the two-layer framework are openly available in the GitHub repository at https://github.com/zhuangcg/framework-integrating-STF-clustering-and-DTs.

**Fig.4** Schematic diagram of two-layer framework (the output of each step is the input of the next step).

### 3.2.1 Layer 1: Spatiotemporal flow clustering - individual cycling flows clustering

In this paper, the methods in Layer 1 can be divided into three essential steps: *Individual Spatial flow clustering*, *Individual Spatiotemporal flow clustering*, and *Neighbor Individual Spatiotemporal flows (ISTFCs) merging*.

*Individual spatial flow clustering:*

This step aims to extract the daily trajectories of individual bike-sharing users from the spatial perspective. In this study, we apply the spatial flow clustering method proposed by (X. Gao et al., 2020) and make enhancements based on the travel distance of bike-sharing. In the original method, spatial dissimilarity $SD_{ij}$ is the key indicator for clustering, which is calculated as follows:

$$SD_{ij} = \sqrt{sd_{ijo}^2 + sd_{ijd}^2} \qquad [1]$$

where $sd_{ijo}$ and $sd_{ijd}$ respectively represent the spatial dissimilarity between the OD of flows $i$ and $j$, which are defined as follows:

$$\begin{cases} sd_{ijo} = \dfrac{dist(O_i, O_j)}{\alpha \times \min(len_i, len_j)} \\ sd_{ijd} = \dfrac{dist(D_i, D_j)}{\alpha \times \min(len_i, len_j)} \end{cases}, \alpha \times \min(len_i, len_j) \leq 200 \qquad [2]$$

where $dist(O_i, O_j)$ and $dist(D_i, D_j)$ denote the Euclidean distance between the same endpoints of two flows. $len_i$ and $len_j$ are the lengths of two flows, respectively. $\alpha$ is a size coefficient which sets the radius of the boundary circle together with $\min(len_i, len_j)$, as displayed in Fig.4(a). In this paper, referring to existing research (X. Gao et al., 2020; Y. Liu et al., 2022), $\alpha$ is set to 0.3.

However, note that the formula of $SD_{ij}$ determines that the radius of the boundary circle increases with the lengths of the flow, thereby reducing the spatial constraint on flow clustering. Although this feature has limited impact on regional-level flow studies, for individual-level related studies, it introduces the noise into clustering results and increases the uncertainty into the extent of individual's daily activities. For example, when $min(len_i, len_j)$=3000 m, the boundary circle radius is 900 m, covering an area of 2.54 km². Hence, to obtain more realistic individual biking flows, we cap the maximum radius of boundary circle at 200 m. This threshold is considered to better aggregate bike-sharing flows and infer the travel activities of cycling users (Yang et al., 2019; S. Li et al., 2021). After settings parameters, flows $i$ and $j$ are deemed spatially similar if $SD_{ij} \leq 1$ (X. Gao et al., 2020).

Then, taking all riding records of each active bike-sharing user as input, spatial flow clustering is performed according to the algorithm proposed by X. Gao et al. (2020). Finally, each individual spatial flow cluster (ISFC) can be denoted as $\{ID_U, ID_{ISFC}, (O, D), n\}$, where $ID_U$ and $ID_{ISFC}$ are the unique identifiers of a user and his/her each ISFC, $(O, D)$ are the OD medoids of all biking flows in ISFC, and $n$ is the number of flows (i.e., trip records) in ISFC. Notably, given that some ISFCs may include insufficient flows to represent a user's daily mobility, we set a minimum threshold for the flows number in each user's ISFCs through the knee-point detection method (See Appendix A): one-fifth of the number of weekdays with recorded bike-sharing trips. Only the ISFCs that satisfy the threshold requirement are deemed reliable and advance to the next step.

*Individual spatiotemporal flow clustering:*

Based on the results of the spatial flow clustering approach, this step further improves the spatiotemporal flow clustering method proposed by X. Yao et al. (2018) to extract individual daily mobility patterns from the temporal perspective. The core of their method is the measurement of temporal similarity $ts_{ij}$, which is defined as follows:

$$ts_{ij} = \frac{T_i \cap T_j}{T_i \cup T_j} \quad [3]$$

where $T_i = [ot_i, dt_i]$ and $T_j = [ot_j, dt_j]$ denote the time spans of flows $i$ and $j$ in the same ISFC, respectively. $T_i \cap T_j$ is their intersection, while $T_i \cup T_j$ is their union (Fig.4(b)). If the time spans of $i$ and $j$ overlap, $ts_{ij}$ is greater than zero. For instance, when $T_i = [8:00, 8:40]$ and $T_j = [8:15, 8:50]$, $T_i \cap T_j$ is 25 min while $T_i \cup T_j$ is 50 min, then $ts_{ij}$ is 0.5.

It is noteworthy that, due to the individual-level focus and the average of 3.6 bike-sharing trips per weekday among active users, our study deems it is impractical to calculate $ts_{ij}$ for travel flows on specific adjacent dates, as conducted by X. Yao et al. (2018). Instead, this paper concentrates on the temporal distribution of cycling activities within a 24-hour timeframe. Simultaneously, this strategy is also more conducive to capturing the genuine mobility of bike-sharing users, because most residents follow regular daily travel patterns, especially commuting trips. For instance, suppose that $T_i$ above occurs on Monday and $T_j$ on Friday, we still assume that their time spans overlap. Moreover, previous research has validated the application of the temporal similarity indicator in taxi trip data (X. Yao et al., 2018). Nevertheless, bike-sharing trips are typically shorter (the average cycling duration for the dataset we used is around 10 min), which can result in a zero temporal similarity even if the travel time of the two biking flows are sufficiently close (e.g., when $T_i$=[8:05,8:15] and $T_j$=[8:15,8:25], $ts_{ij}$=0). To address this, we introduce an expansion coefficient $\beta$ to $T_i$ and $T_j$ (i.e., $T_i = [ot_i - \beta, dt_i + \beta]$ and $T_j = [ot_j - \beta, dt_j + \beta]$) to ensure that the time-adjacent cycling flows can be identified and clustered. In this study, $\beta$ is set to 30 min (more details in Section 4.1). After the above improvement, referring to the original method, we consider that the travel times of flows $i$ and $j$ are adjacent when $ts_{ij} \geq 0.5$.

Later, we use the biking records including in each user's ISTC as input and execute the spatiotemporal flow clustering algorithm by X. Yao et al. (2018). Each ISTFC can be denoted as $\{ID_{USER}, ID_{ISFC}, ID_{ISTFC}, (O, D), n', T_o, T_d\}$, where $ID_{ISFC}$ is the unique identifier of the ISFC to which the ISTFC belongs, $ID_{ISTFC}$ is the unique identifier of each ISTFC. $n'$ is the flows number in the ISTFC, and $T_o$ and $T_d$ are the average starting and ending time of the biking flows in each ISTFC, respectively. The resulting ISTFCs are used for next processing.

*Neighbor ISTFCs merging*:

By observing the result of "***Individual spatiotemporal flow clustering***", we find that some ISTFCs are spatiotemporally adjacent but not merged, as illustrated in Fig.4(c). The reasons are relevant to two aspects: First, some bike-sharing users have multiple optional routes to and from the same daily activity places. The locations of ODs (e.g., different entrances to an industrial park) and the direction of their trip flows vary with the different routes, which leads to difficulties in clustering them into the same ISTFC. Second, in "***Individual spatial flow clustering***" step, the restriction of boundary circle may result in dividing the cycling flows into more ISFCs. Nevertheless, the improvement of $SD_{ij}$ is essential to extract more accurate individual daily trajectories. To improve the utilization of biking records for these affected users, we examine and merge neighbor ISTFCs in the last step of Layer 1.

Given a set of all ISTFCs for an active bike-sharing user $FC$ and the size coefficient $\alpha$, the process of merging neighbor ISTFCs is shown in Algorithm 1. In short, two ISTFCs $FC_i$ and $FC_j$ that can be merged must satisfy the following conditions:

(1) The temporal similarity $ts_{ij}$ is not less than 0.5, which is consistent with "***Individual spatiotemporal flow clustering***" step;

(2) The boundary circle at the same endpoints of $FC_i$ and $FC_j$ must intersect (i.e., the distance between these endpoints should be less than twice the radius of the boundary circle $r$, which is calculated consistent with "***Individual spatial flow clustering***" step).

When $FC_i$ and $FC_j$ satisfy the above conditions, $FC_j$ is merged by $FC_i$.

---

**Algorithm 1** Merging Neighbor ISTFCs

---

**Input:** $FC = \{FC_i | 1 \leq i \leq n\} \leftarrow$ a set of all ISTFCs for an active bike-sharing user, in descending order according to the flows number they contain; and $\alpha \leftarrow$ the size coefficient;

**Steps: For each** ISTFC $FC_i$, where $1 \leq i \leq n$

**For each** ISTFC $FC_j$, where $i < j \leq n$

$r = \alpha \times min(len_i, len_j)$

**If** $r > 200$ **then** $r = 200$

**If** $ts_{ij} \geq 0.5$ and $dist(O_i, O_j) < 2r$ and $dist(D_i, D_j) < 2r$ **then**

Merge the two ISTFCs: $FC_i \leftarrow FC_i \cup FC_j$ and $FC \leftarrow FC/FC_j$

**Return:** A set of all ISTFCs for this user after merging $FC = \{FC_i | 1 \leq i \leq m\}$.

---

It is worth noting that the inputted ISTFCs for each user should be sorted in descending order based on the flows number they contain. This is because the ISTFCs with more biking records are more likely to represent the most typical daily mobility patterns of a user, and prioritizing these ISTFCs during the traversal process ensures that these predominant patterns are retained after merging processing. In addition, through the above ISTFCs merging process, some ISTFCs still have limitations in capturing the individual daily mobility due to the small number of trips they contain. To do so, we also set a minimum threshold for filtering ISTFCs through the knee-point detection method (See Appendix A): an ISTFC must contain at least 30% of the number of biking flows within its corresponding ISFC (i.e., $n' \geq 0.3 \times n$). Only the ISTFCs that satisfy this threshold are deemed reliable and employed as inputs to Layer 2.

**3.2.2 Layer 2: Rule-based decision trees - daily cycling commuting behaviors mining**

While the ISTFCs acquired in Layer 1 capture the spatiotemporal patterns of individual daily mobility, they lack semantic information about the travel activities. In Layer 2, we build three rule-based decision trees considering round-trip journeys, working hours, and public transport transfers, aiming to mine individual commuting patterns from the ISTFCs of bike-sharing users.

Initially, we develop the "***Candidate commuting flow identifier***" decision tree (Fig.5), focusing on round-trip frequencies and working hours. This identifier aims to extract the individual candidate commuting flows (ICCFs) from the ISTFCs of each user. Specifically, we define an ICCF by below two criteria:

(1) The commuting behavior should be characterized by frequent and symmetrical (i.e., round-journeys) travel flows between two locations (Y. Liu et al., 2024);

(2) There should be a substantial time interval between the biking flows in opposite directions, symbolizing the user's daily working hours.

Hence, in "***Candidate commuting flow identifier***", given two ISTFCs $FC_i$ and $FC_j$ for a bike-sharing user, we first assess the spatial adjacency of their opposite endpoints. To be specific, we require that both $dist(O_i, D_j)$ and $dist(O_j, D_i)$ to be less than twice of boundary circle radius $r$. The calculation of $r$ is consistent with Layer 1. If $FC_i$ and $FC_j$ meet the spatial adjacency, we further evaluate whether the time interval between them exceeds the minimum working hours threshold $T_{wh}$, as shown in Fig.4(d). Following Sari Aslam et al. (2019), this paper establishes $T_{wh}$ at 4 hours to effectively capturing the daily working behaviors of full-time, part-time, and shift workers. If $FC_i$ and $FC_j$ satisfy the above two conditions, they are marked as an ICCF. Each ICCF can represented as $\{ID_U, ID_{ICCF}, ISTFC_e, ISTFC_l\}$, where $ID_{ICCF}$ is the unique identifier of the ICCF, while $ISTFC_e$ and $ISTFC_l$ denote the ISTFCs with earlier and later travel time, respectively. Notably, we define the STFCs closer to 8 a.m. as $ISTFC_e$, while the other is designated as $ISTFC_l$, as this division aligns with the daily routine of the vast majority of residents. Moreover, before inputting the user's ISTFCs into the identifier, we also sort them by the flows number they encompass in descending order to encapsulate the user's most predominant daily activity patterns.

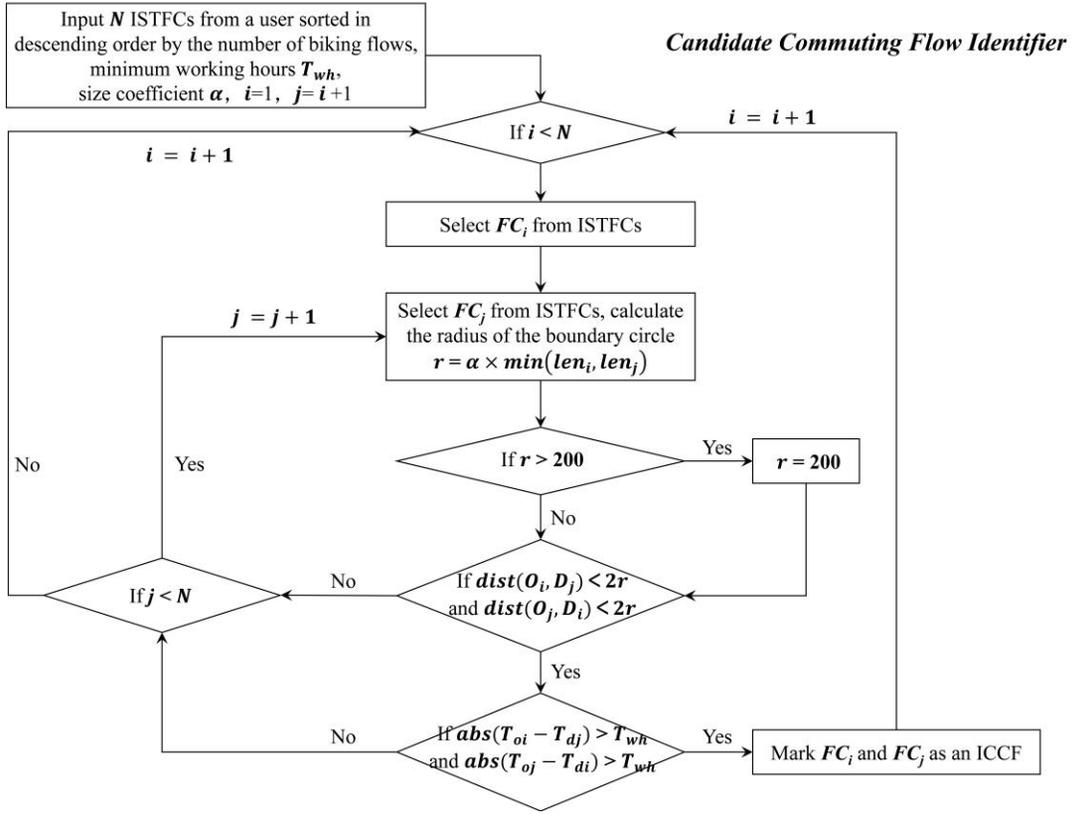

**Fig.5** Flowchart of candidate commuting flow identifier.

Afterwards, to facilitate subsequent analysis, we simplify the identified ICCF consisting of a pair of ISTFCs into a single flow (Fig.4(e)). The direction of simplified ICCF (SICCF) is set to match the ISTFC with earlier travel time (i.e., $ISTFC_e$). Thus, for each SICCF, its origin (i.e., $O_{ICCF}$) is the midpoint of the origin of $ISTFC_e$ and the destination of $ISTFC_l$. Conversely, $D_{ICCF}$ is the midpoint of the origin of $ISTFC_l$ and the destination of $ISTFC_e$. Additionally, we define the following eight attributes for each SICCF to identify and analyze individual daily commuting patterns:

(1) Departure time for the ISTFC with earlier travel time ($T_e$): the $T_o$ of $ISTFC_e$;

(2) Departure time for the ISTFC with later travel time ($T_l$): the $T_o$ of $ISTFC_l$;

(3) Cycling duration for the ISTFC with earlier travel time ($CT_e$): the difference between $T_o$ and $T_d$ of $ISTFC_e$;

(4) Cycling duration for the ISTFC with later travel time ($CT_l$): the difference between $T_o$ and $T_d$ of $ISTFC_l$;

(5) Cycling commuting distance ($CD$): the Euclidean distance between $O_{ICCF}$ and $D_{ICCF}$;

(6) Working hours ($WH$): the difference between the $T_d$ of $ISTFC_e$ and the $T_o$ of $ISTFC_l$;

(7) Total number of biking flows ($n_t$): the sum of the biking flows in $ISTFC_e$ and $ISTFC_l$;

(8) Cycling round-trip rate ($R_{rt}$): the ratio of the number of biking flows in $ISTFC_e$ and $n_t$, this indicator can measure the imbalance in commuting frequencies between the two opposite directions

Each SICCF is noted as $\{ID_U, ID_{ICCF}, (O_{ICCF}, D_{ICCF}), T_e, T_l, CT_e, CT_l, CD, WH, n_t, R_{rt}\}$.

Next, we establish the "***Transfer commuting flow identifier***" decision tree (Fig.6), accounting for public transit transfers, to identify latent daily transfer commuting behaviors from users' SICCFs. This consideration arises from research indicating that transferring to public transportation, especially the metro, is the important travel purpose of bike-sharing (Xing et al., 2020; S. Li et al., 2021). Additionally, the integrated use of bike-sharing and public transport has attracted significant research attention recently (Ma et al., 2018; Guo & He, 2020; Fu et al., 2023; Zhu et al., 2024; Zhang et al., 2024). Therefore, it is crucial to determine whether bike-sharing users regularly ride to connect with public transit for their daily commuting. The workflow of the identifier in Fig. 6 is described as follows:

(1) Take the public transport station data and a user's SICCF as input, and set a maximum transfer distance threshold $TD$. In this study, the $TD$ is set to 60 m for metros referring to S. Liu et al. (2022), and 30 m for buses, which are deemed less attractive for bike-sharing (Guo & He, 2020).

(2) If the SICCF's departure time is outside the public transportation operating hours (from 6:00 to 23:30 in our study area), it is considered not connected to public transport. Conversely, we continue.

(3) Identify the nearest public transport stations to the OD of the SICCF (i.e., $O_{ICCF}$ and $D_{ICCF}$), labeled as $S_o$ and $S_d$, respectively. If $dist(O_{ICCF}, S_o)$ and $dist(D_{ICCF}, S_d)$ both exceed $TD$, this SICCF is deemed to not connected to public transit. Conversely, we proceed.

(4) If $dist(O_{ICCF}, S_o)$ is smaller than $TD$ and $dist(D_{ICCF}, S_d)$, i.e., the origin of SICCF is closer to its nearest transport station, we still cannot conclude that this user regularly rides bike-sharing between the workplace and the transfer station. This is because public transport stations often coexist with various activity places, especially around metro stations (S. Liu et al., 2022). In this study, we need further compare the distance from the destination of SICCF (i.e., $D_{ICCF}$) to its nearest transport station (i.e., $S_d$) with the length of this SICCF (i.e., $CD$). If $2 \times dist(D_{ICCF}, S_d) \leq CD$, it is argued that the SICCF is not connected to public transit, as the user has chosen a longer cycling route instead of a shorter journey from $S_d$ to $D_{ICCF}$ (see Fig.B.1 in Appendix B). Conversely, it is inferred that the origin of this SICCF is connected to public transport. Similarly, if $dist(D_{ICCF}, S_d)$ is smaller than $TD$ and $dist(O_{ICCF}, S_o)$, we employ the same method to determine whether the destination of this SICCF is connected to public transport (see Fig.4(f)).

Note that for each SICCF, we employ the "**Transfer commuting flow identifier**" to assess connections with bus and the metro systems. When a SICCF qualifies for connectivity with both, the metro is prioritized over the bus (Guo & He, 2021).

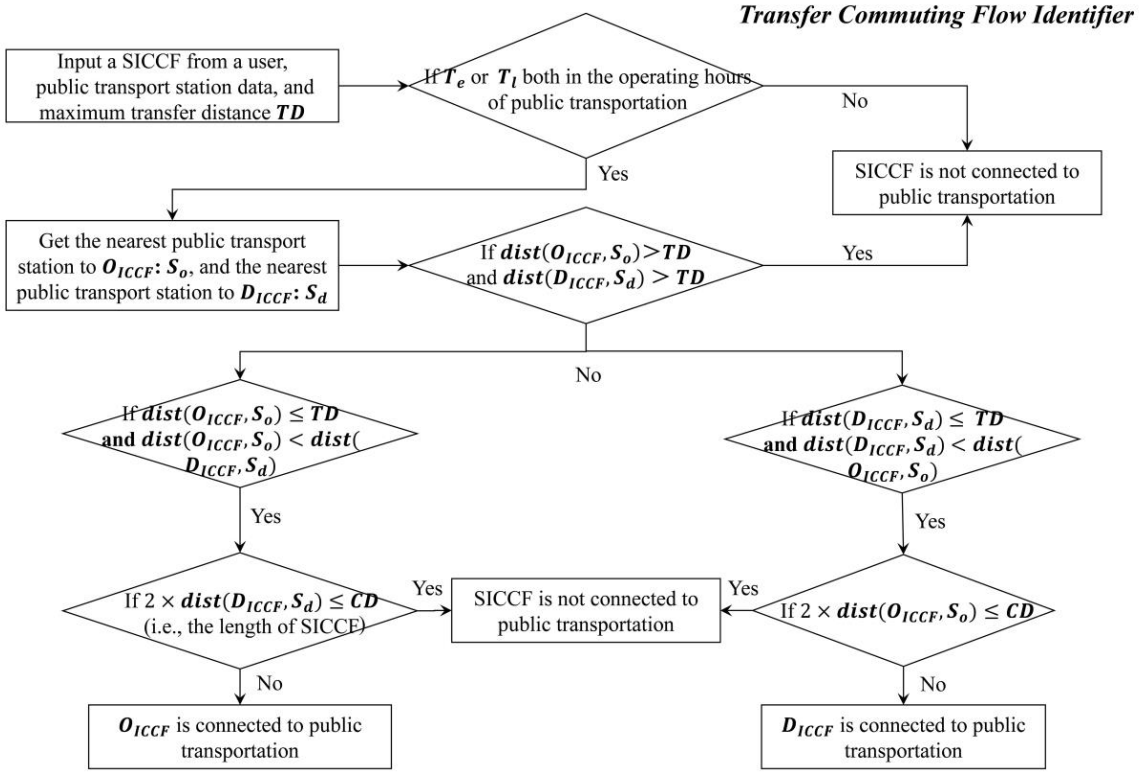

**Fig.6** Flowchart of transfer commuting flow identifier.

Finally, we build the "**Biking commuting user classifier**" decision tree to categorize the most predominant daily commuting patterns among individual bike-sharing users (Fig.7). In our study, the SICCF contained the highest number of biking flows (i.e., $n_t$) for each user is deemed most representative of his/her daily commuting patterns during the study period, which is designated as the individual daily commuting flow (IDCF). Based on the IDCFs of bike-sharing users, we classified them into two main categories: *Only-biking* and *Biking-with-transit commuters*, and the latter is further subcategorized into: *Biking-transit*, *Transit-* *biking*, and *Biking-transit-biking commuters*, drawing insights from relevant studies (Patrick A. Singleton & Kelly J. Clifton, 2014; Guo et al., 2021). The definition of each user category in Fig.7 are outlined as follows:

(1) Take the IDCF for a user as the input.

(2) If the IDCF lacks a connection to public transit, this user is classified as an *Only-biking commuter* who relies solely on biking for his/her daily home-work commuting. The OD of the IDCF are his/her residence and workplace, respectively. Conversely, it proceeds to the next step.

(3) If the origin of the IDCF is connected to public transit, it signifies that this IDCF represents the user's daily "last-mile" commuting to work by bicycling from a transit station (or the "first-mile" biking from his/her workplace to the transit station after work). The origin of this IDCF indicates the transit station where the user starts daily his/her cycling to work, while the destination is his/her workplace. However, in this scenario, the daily commuting chain of this user is incomplete, as it lacks the segment where the user travels between the home and the transfer station. Hence, we need search for his/her remaining SICCFs that satisfy the following conditions to form his/her complete daily commuting chain:

- The destination of this SICCF is connected to public transit;
- The transfer station of this SICCF and the IDCF are different;
- This SICFF is temporal close to the IDCF, meaning the time difference between this SICCF's and the IDCF's $T_e$, minus the $CT_e$ of SICCF, or the time difference between this SICCF's and the IDCF's $T_l$, minus the $CT_l$ of IDCF, is less than 2 hours. This threshold is chosen to take into account the relatively longer duration of metro and bus travel, as well as the potential variability in residents' routine travel schedules.

If an SICCF meeting the above conditions is found, it is labeled as an individual additional daily commuting flow (IADCF), and the process proceeds to the next step. Otherwise, the user is considered a *Transit-biking commuter*, who relies solely on biking for the "last mile" from transit station to his/her workplace (or the "first mile" from his/her workplace to transit station after work). Similarly, if the destination of a user's IDCF is connected to public transit but no suitable IADCF (its origin connect to another transport station) is identified among the remaining SICCFs of this user, his/her is categorized as a *Biking-transit commuter*, who relies exclusively on bicycling for the "first mile" from his/her residence to transit station (or the "last mile" from transit station to his/her residence after work).

(4) If the origin of the user's IDCF is connected to a transit station and an IADCF is identified, this IADCF represents the user's daily "first mile" commuting by bicycling from his/her residence to another transit station (or the "last mile" commuting when returning home from another transport station after work). In this scenario, by combining the IDCF with the IADCF, the complete daily commuting pattern, including home and work locations, can be established. Meanwhile, this user is categorized as a *Biking-transit-biking commuter* (See Fig.B.2 in Appendix B). Likewise, if the destination of a user's IDCF is connected to public transit while an IADCF is found, this user is also classified as a *Biking-transit-biking commuter*.

In Fig.8(a), we illustrate a schematic diagram of each category of cycling commuters.

Moreover, there are differences in the commuting characteristics among various categories of commuters. For more details, please refer to Appendix C in Supplemental files.

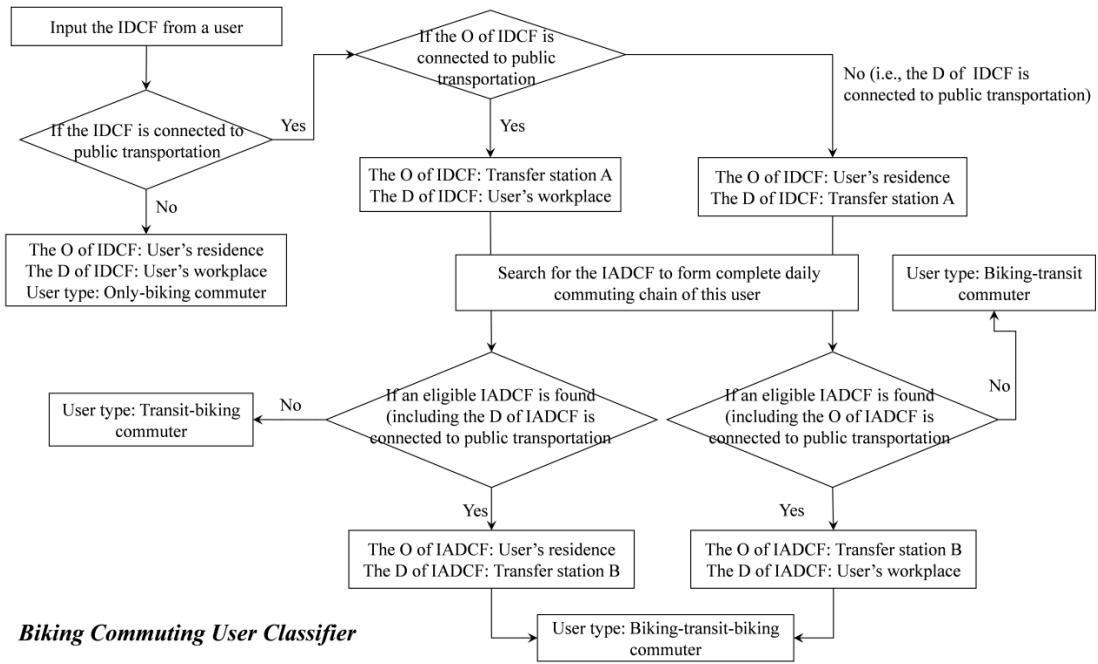

**Fig.7** Flowchart of biking commuting user classifier.

*3.3 Framework evaluation and validation*

To demonstrate the feasibility and applicability of our proposed two-layer framework, this paper evaluate the performance of the improved spatiotemporal flow clustering method through comparing the original method and validate the identification results of individual residential locations of bike-sharing commuters through spatial proximity analysis.

For Layer 1, we contrast the clustering results of the original methods with our enhanced methods using multiple indicators. For ***Individual spatial flow clustering*** method, we computed four indicators: the average number of biking records included in all ISFCs, the average length of all ISFCs, and the average distance from the OD of each biking record to its corresponding ISFC's OD in all ISFCs (later abbreviated as the average distances to ISFCs' ODs, respectively). These indicators are used to highlight the promoting of restricting the boundary circle for more precise daily cycling trajectories extraction. For ***Individual spatiotemporal flow clustering*** method, we examine the impact of different expansion coefficients $\beta$ on the average number of biking flows contained in all ISTFCs and the average maximum time interval for all ISTFCs (i.e., the average of the difference between the earliest departure and the latest arrival times of the trip records for all ISTFCs). This analysis aims to illustrate the necessity of expansion coefficient in mining spatiotemporal mobility for bike-sharing trips.

For Layer 2, due to the unavailability of travel survey data on cycling habits within the study area and the large discrepancy between the census population and the bike-sharing groups, we decide to employ actual residential land use data to validate the accuracy of our extracted users' home locations. Specifically, we first extract the users who have identifiable residential locations, then measure the distances from their residence to the nearest actual boundaries of residential land parcels. If a user's residence falls in the residential land, the distance is set to zero. Lastly, we plot the cumulative percentage of users whose identified residences are within 0 to 300 meters of the actual residential land. If the majority of users' residences are located within or near the actual residential land parcels, it would demonstrate a close match between the bike-sharing commuters' residences and the actual residential land use distribution. This serves as a proxy to demonstrate the validity of the individual daily commuting patterns identified in this paper. Notably, we do not perform the same validation for users' workplaces identified, as work locations for residents with different occupations vary widely and are not limited to office buildings or industrial parks. This diversity increases the likelihood of omissions and misclassifications.

*3.4 Aggregation and visualization analysis*

Based on the identification and validation results of bike-sharing commuters, we further aggregate and analyze their daily commuting characteristics (i.e., commuting duration and distance, working hours, and cycling round-trips rate) and spatiotemporal patterns (i.e., commuting temporal patterns, spatial distribution of residences, workplaces, metro transfer stations, and daily commuting chains) within the study area.

**4 Result and discussion**

*4.1 Comparative analysis and validation results*

In Table 2, we compare the evaluation indicators between the original and improved spatial flow clustering methods. Obviously, in comparison with the improved method, the original method exhibits a slight increase in the average number of trip records in all ISFCs (an average increase of 3 additional records) due to the absence of the boundary circle constraints. Meanwhile, notable changes are observed in the average distances to ISFCs' ODs, with increases of 23 m and 16 m, respectively.

Considering the significant impact of the boundary circle radius constraint on longer ISFCs, we conducted additional comparisons for ISFCs exceeding lengths of 1500 m and 3000 m. The results reveal that in the enhanced method, the average distances to the ISFCs' ODs maintains nearly constant as ISFC length increasing, whereas in the original method, it increases substantially. However, the amplification of these two indicators implies significant uncertainty in determining the OD of longer ISFCs, as the coverage of their boundary circles is excessively broad. These uncertainties may introduce more inaccuracies in subsequent analyses (e.g., identifying users' residences and workplaces). Consequently, the enhancement of *Individual spatial flow clustering* method introduced in this study are essential, ultimately extracting reliable ISFCs from 95.1% (~0.71 million) of active bike-sharing users.

**Table 2** Comparison of evaluation indicators between the original and improved spatial flow clustering methods (see Section 3.3 for descriptions of below indicators).

| Method | Original method | Improved method |
| --- | --- | --- |
| **Avg. number of biking records** | 43.5 (26.7) | 40.7 (25.3) |
| **Avg. distance to ISFCs' origins (unit: m)** | 105 (111) | 82 (64) |
| **Avg. distance to ISFCs' destinations (unit: m)** | 80 (89) | 64 (51) |
| **Pct. of ISFCs more than 1500 m** | 24.40% | 23.06% |
| **Avg. distance to ISFCs' origins (> 1500 m)** | 162 (161) | 83 (63) |
| **Avg. distance to ISFC's destinations (> 1500 m)** | 118 (138) | 63 (50) |
| **Pct. of ISFCs more than 3000 m** | 4.69% | 4.21% |
| **Avg. distance to ISFCs' origins (> 3000 m)** | 214 (232) | 84 (64) |
| **Avg. distance to ISFCs' destinations (> 3000 m)** | 156 (215) | 61 (49) |

\* The values in bracket are standard deviations of the corresponding indicators.

Similarly, Table 3 displays the comparative results of the original and improved spatiotemporal flow clustering methods. It is clearly that in contrast to the original method ($\beta=0$), the improved method, incorporating an expansion coefficient $\beta$, can extract ISTFCs that contain more trip records (averaging an increase of 7.1 records including in each ISTFC when $\beta=30$ min). This demonstrates the contribution of the $\beta$ in mining daily spatiotemporal trajectories from bike-sharing data, given the generally shorter travel durations for bicycle trips. However, as $\beta$ increases further, the average number of trip records within ISTFCs shows diminishing returns, with an increase of only 2.8 records at $\beta=90$ min compared to $\beta=30$ min. Meanwhile, the average maximum time interval for all ITSFC continues to increase with the growing of $\beta$. Yet, an excessively large average maximum time interval could introduce biking records from other time periods into the extracted ISTFCs, potentially elevating data noise. Hence, this study refers to the China Urban Transportation Report 2021 (https://jiaotong.baidu.com/cms/reports/traffic/2021/index.html) and select a final $\beta$ value of 30 min, which is remarkably close to the average commuting duration in Shenzhen (37 min). Ultimately, through the steps of *Individual spatiotemporal flow clustering* and *Neighbor ISTFC merging*, we successfully identify reliable ISTFCs from 74.4% (~0.56 million) of active bike-sharing users. Notably, ~0.11 million reliable ISTFCs from over 90,000 users are accomplished through *Neighbor ISTFC merging* step. Collectively, these results underscore the critical role of the aforementioned enhancements in Layer 1 in improving the quality of daily travel trajectories extraction for bike-sharing users.

**Table 3** Comparison of evaluation indicators between the original and improved spatiotemporal flow clustering methods (see Section 3.3 for descriptions of below indicators)

| **Method** | **Original method** | **Improved method** | | |
|---|---|---|---|---|
| | $\beta=0$ | $\beta=30$min | $\beta=60$min | $\beta=90$min |
| **Avg. number of biking records** | 12.5 (10.3) | 19.6 (14.8) | 21.3 (15.3) | 22.4 (15.6) |
| **Avg. maximum time interval for all ITSFCs (unit: min)** | 16.8 (11.7) | 34.8 (13.9) | 49.1 (20.3) | 61.8 (27.0) |

\* The values in bracket are standard deviations of the corresponding indicators.

Furthermore, utilizing the rule-base decision trees from Layer 2, we have extracted the IDCFs of 383,786 active bike-sharing users with reliable ISTFCs. Fig.8(b) illustrates the proportion of identified bike-sharing commuters in different categories: 74.38% are *Only-biking commuters* and 25.62% are *Biking-with-transit commuters*. The percentage of *Biking-with-transit commuters* is slightly higher than the results for transfer trips in the studies of Xing et al. (2020) and S. Li et al. (2021) regarding the purpose of bike-sharing trips, while they considered more kinds of travel activities. Within these *Biking-with-transit commuters*, the share of *Biking-transit-biking commuters* is only 1.75% due to the stringent filtering rules, while *Biking-transit commuters* (14.39%) are more prevalent than *Transit-biking commuters* (9.48%), aligning with the findings of Guo et al. (2021), which suggests that more users rely on cycling for the "first mile" from residence to transit station (or the "last mile" from transit station to home after work). Meanwhile, given that most bike-sharing commuters daily transfer to the metros (over 96%) rather than the buses, our subsequent analysis will focus on the integrated biking-metro commuting patterns.

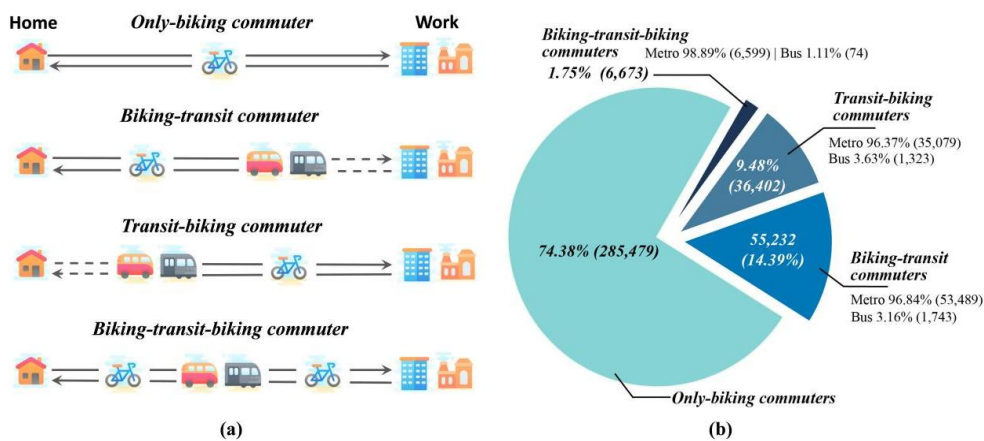

**Fig.8** Schematic diagram (a) and percentage (b) of different categories of bike-sharing commuters.

To further validate the accuracy of the identification results of bike-sharing commuters, we measure the proximity between the their identified residences and the actual residential land use boundaries (see Section 3.3 for details), as showed in Fig.9. The result illustrates that 51.5% of inferred users' residences are within the residential land parcels, and 93.5% are within 100 m of the residential land use. This indicates that the most of the identified users' residences are adjacent to the actual residential land boundaries, reflecting the feasibility of our two-layer framework.

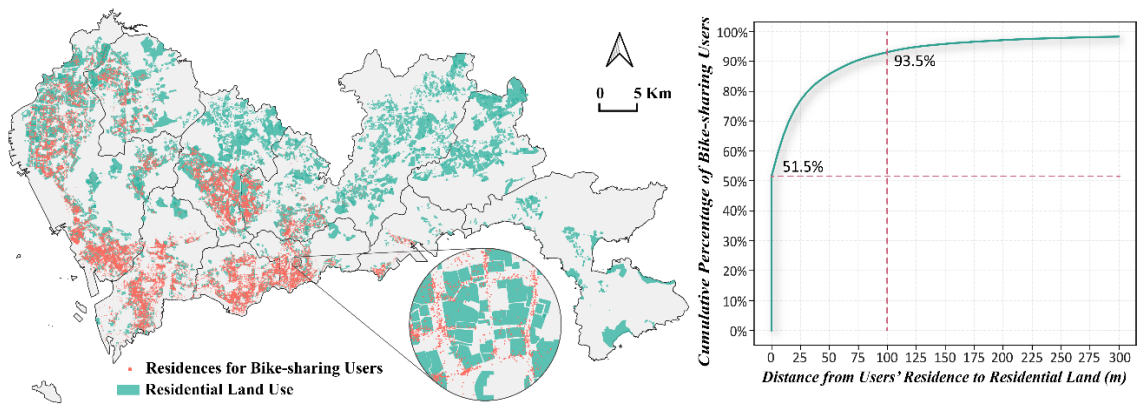

**Fig 9.** Cumulative percentage of bike-sharing users whose identified residences fall within the residential land boundaries.

*4.2 Commuting characteristics among bike-sharing users*

### 4.2.1 Commuting duration and distance

Fig.10 shows the distribution of commuting duration and distance for *Only-biking* and *Biking-transit-biking commuters*. Since the commuting chains for *Transit-biking* or *Biking-transit commuters* are incomplete, we cannot discuss these commuting characteristics for them. For *Only-biking commuters* (Fig.10(a, b)), we find that over three-quarters have a daily commuting duration under 10 min and distance within 1.8 km, aligning with previous studies on the characteristics of bike-sharing usage (Shen et al., 2018; Ma et al., 2020; F. Gao et al., 2022). This result suggests that most *Only-biking commuters* are more likely to reside near their workplaces. For *Biking-transit-biking commuters* (Fig.10(c, d)), we observe an average commuting duration exceeding 45 min and distance over 13 km, indicating that these users may tend towards complete their daily home-work commuting across districts. Moreover, when comparing the commuting duration distribution for different trip purposes (Fig.10(a, c)), we discover that both categories of bike-sharing commuters spend more time commuting home from work. The findings is consistent with the research of Kung et al. (2014), which can be attributed to having more intervening opportunities for other activities (e.g., recreation, shopping and etc.) during their journey home (after work).

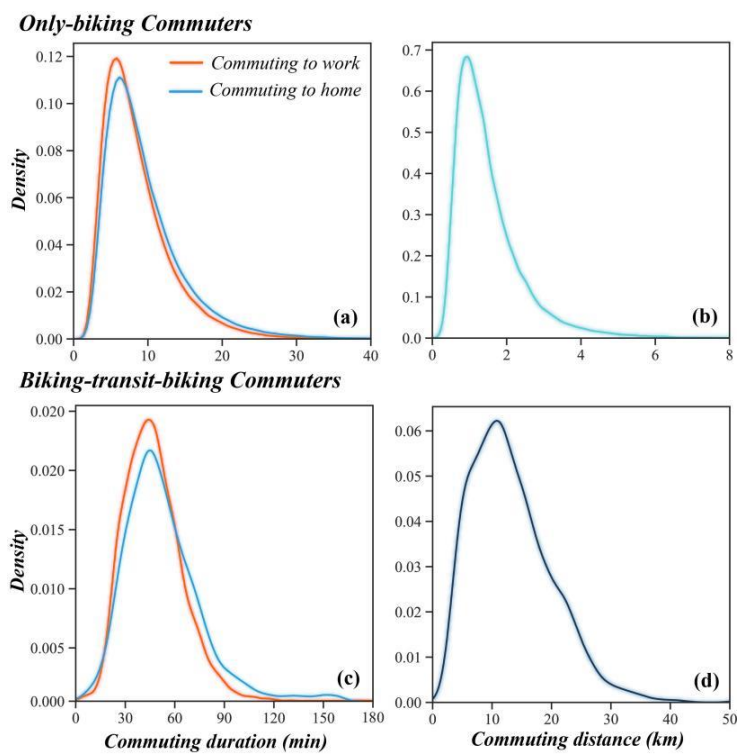

**Fig. 10** Distributions of commuting duration (a, c) and commuting distance (b, d) for *Only-biking* and *biking-transit-biking commuters*

**4.2.1 Working hours**

Fig.11(a) displays the distribution of working hours for all bike-sharing commuters, excluding *Biking-transit commuters*, as we cannot obtain their daily arrival and departure times at their workplaces. Specifically, we observe three distinct peaks in the working hours of *Only-biking commuters*. The largest peak occurs around 10 hours, which is longer than the typical 8-hour workday. However, note that the working hours we calculated represent the total time from user's daily arrival at the workplace to their departure, potentially including non-working hours like lunch breaks. Thus, the actual working hours for many individuals may be 1 to 2 hours shorter than the working hours we calculated, indicating that the working hours for most users are in accordance with legal regulations. The second highest peak appears approximately 12.5 hours, suggesting that some users are actually working overtime, even if their calculated working hours include break times. Lastly, there is a smallest peak around 5 hours, significantly lower than the first two peaks, which represents a minority of individuals working part-time or on shift.

In comparison to *Only-biking commuters*, *Biking-transit* and *Biking-transit-biking commuters* exhibit a single prominent peak in their working hours, which aligns with the largest peak for *Only-biking commuters*. Furthermore, although some *Biking-with-transit commuters* also work overtime, as indicated by a slight peak after 12 hours, this proportion is notably lower than *Only-biking commuters*. This reflects that *Only-biking commuters* are more tolerant of overtime compared to *Biking-with-transit commuters*, potentially due to their lower commuting costs. Lastly, we discover that bike-sharing users who involved in part-time or shift work rarely use public transport for commuting. That is reasonable, as their working hour are around 5 hours, and choosing a "Biking-transit" commuting mode would represents excessively high proportion of their overall commuting duration relative to their working hours (Schwanen & Dijst, 2002).

### 4.3.3 Cycling round-trip rate

Cycling round-trip rate serves as an indicator of the regularity for cycling commuting trips, capturing differences in cycling habits between trips to and from work. Generally, as show in Fig.11(b), there is little difference in the cycling round-trip rates among various kinds of commuters (note that *Biking-transit-biking commuters* are fewer and thus have a more concentrated distribution). The average cycling round-trip rate is around 0.6, with the lower quartile roughly 0.5, indicating that for nearly three-quarters of bike-sharing commuters, riding to work is more regular than riding home. In other words, most users prefer to bike to work rather than from work. This could be due to fewer time constraints and more flexibility activities for residents after work. Additionally, it may reflect the insufficient supply of bike-sharing during the off-duty peak hours, prompting some users to choose alternatives for their return trip. For bike-sharing operators, it is crucial to address this issue by increasing the supply of bike-sharing in hotspot areas during the evening peak, based on long-term usage patterns, which could help enhance user retention and loyalty.

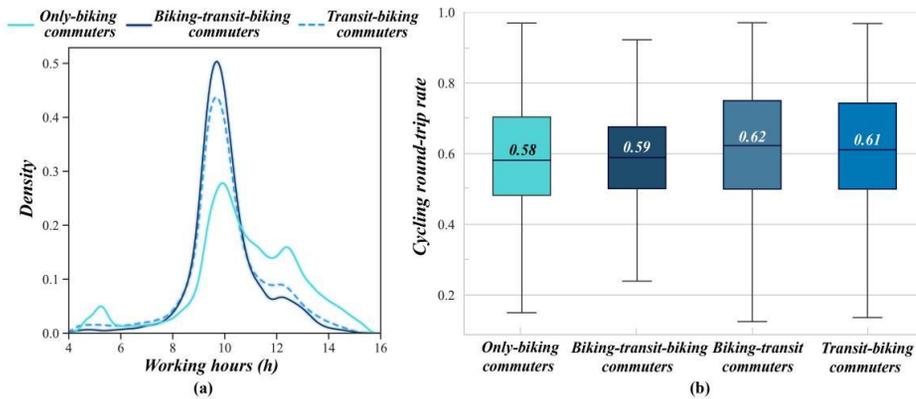

**Fig. 11** (a) Distribution of working hours for *Only-biking*, *Transit-biking* and *Biking-transit-biking commuters*; (b) Distribution of cycling round-trip rate for different categories of bike-sharing commuters.

*4.3 Commuting spatiotemporal patterns among bike-sharing users*

### 4.3.1 Temporal patterns of bike-sharing commuters

In Fig.12, we present the daily commuting temporal patterns for different kinds of bike-sharing commuters. However, due to the incomplete commuting chains of *Transit-biking commuters*, as mentioned earlier, we cannot discuss their home-to-work temporal patterns. Similarly, we also omit the work-back-home temporal patterns for *Biking-transit commuters*.

Regarding the home-to-work temporal patterns (Fig.12(a)), we observe that the peak departure times for *Biking-transit* and *Biking-transit-biking commuters* both occur before 8:00, while the peak for *Only-biking commuters* appears around 8:30. Combined with the observation from Fig.10, this suggests that users with higher commuting costs tend to depart earlier, in line with the findings by Kung et al. (2014). Moreover, *Biking-transit commuters* tend to depart slightly later than *Biking-transit-biking commuters*, indicating that their overall commuting durations may be shorter, with workplaces located closer to the transit stations.

With regard to the work-back-home temporal patterns (Fig.12(b)), we find that the peak of all three kinds of commuters occurs around 18:30, which reflects the standard off-duty commuting time for most bike-sharing users. However, this also means a massive demand for bikes during this period, particularly around the workplaces. If bicycles availability is insufficient, some users are forced to choose alternative transportation, which helps explain why the cycling round-trip rate for most users are greater than 0.5 (Fig.11(b)). Furthermore, compared to home-to-work pattern, the work-back-home curve is smoother, with a longer tail (20:00-23:00), again reflecting the phenomenon of overtime for some users, especially among *Only-biking commuters*. This result echoes the discussion in Fig.11(a). Meanwhile, this observation also highlights that the substantial commuting demand during evening hours, particularly for *Only-biking commuters*. Correspondingly, urban planners should prioritize improving lighting infrastructure in areas with high night-time bike-sharing usage to enhance the safety and satisfy commuters needs during these hours.

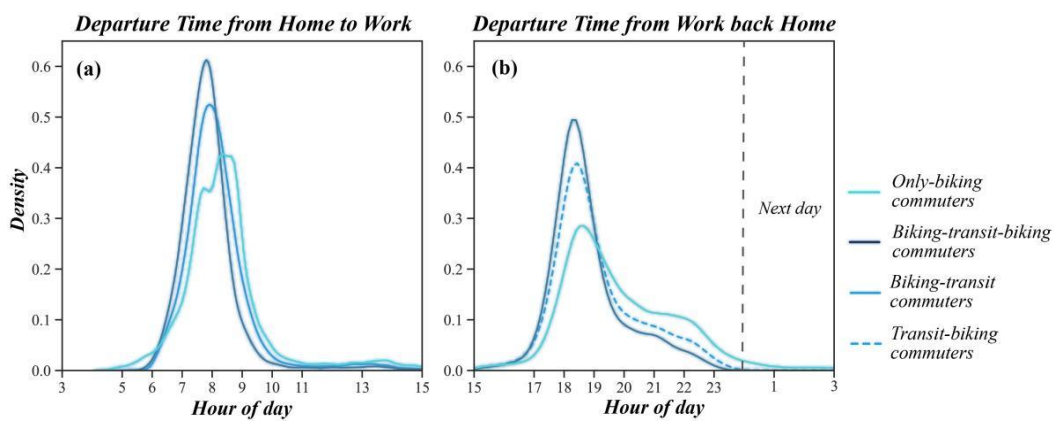

**Fig. 12** (a) Daily temporal patterns of commuting to work for *Only-biking*, *Biking-transit-biking* and *Biking-transit commuters*; (d) Daily temporal patterns of commuting home for *Only-biking*, *Biking-transit-biking* and *Transit-biking commuters*

**4.3.2 Spatial distribution of workplaces and residences for bike-sharing commuters**

Fig.13(a) illustrates the density distribution of residential locations for all bike-sharing commuters, excluding *Transit-biking commuters* whose residences cannot be identified. Likewise, Fig.13(b) shows the distribution of workplace for all commuters except for *Biking-transit commuters*. Generally, the spatial distribution of residential and employment area for bike-sharing commuters exhibits both widespread dispersion and local concentrations. Specifically, employment hotspots are predominantly in areas such as the Futian FTZ – Futian CBD – Luohu CBD, High-tech Park – Bao'an Center and Longhua Industrial Park, with most residential hotspots distributed nearby. This result is in line with the mixed-use land patterns in Shenzhen. Interestingly, we find that the main residential hotspots are concentrated in urban villages and old communities, especially in the central city. Due to lower living costs, these areas attract large numbers of young migrants and recent graduates seeking rental housing (Y. Liu et al., 2010). Concurrently, Guo et al. (2021) noted that this demographic forms the main force of bike-sharing users in Shenzhen. Moreover, the narrow roads, high-density buildings, and mixed land use in these areas further support flexible and convenient bicycle trips. Therefore, despite the operational challenges of managing and reallocating bikes within these complex urban villages and old communities, these areas have substantial mobility demand (especially for commuting) that warrants the attention of bike-sharing operators.

Furthermore, we calculate the average working hours at the major employment centers in Shenzhen. Notably, as discussed in Subsection 4.2.1, the computed working hours are longer than actual working hours for most users; however, this discrepancy does not impede inter-regional comparison. The result shows that average working hours in central city areas are generally shorter in the suburban areas (Fig.13(b)). Specifically, in the central city, employment centers dominated by commercial and service industries (e.g., Luohu CBD and Futian CBD) exhibit shorter working hours than those centered on high-tech industries (e.g., High-tech Park), and the Huaqiang North Commercial Area in Futian District has the shortest average working hours (9.89h). In contrast, in the suburbs, the Longhua Industrial Park, predominantly manufacturing-based, has the longest average working hours (10.89h), which suggest a higher likelihood of overtime for bike-sharing users employed there. Accordingly, bike-sharing operators could devise targeted bike reallocation strategies tailored to varying off-duty hours at different employment centers, thereby improving the utilization of bike-sharing across regions.

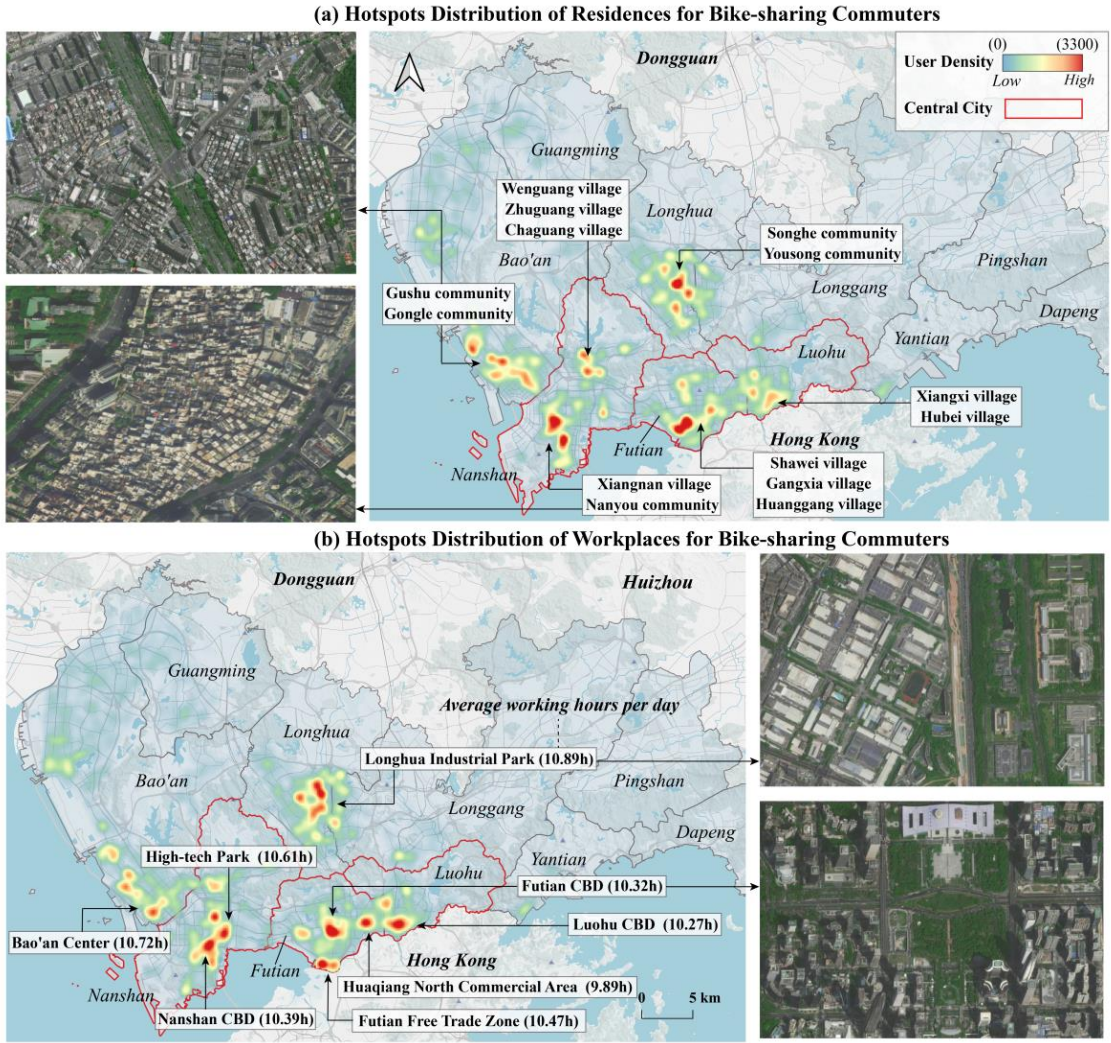

**Fig. 13** Hotspots distribution of residences and workplaces for bike-sharing commuters

### 4.3.3 Spatial distribution of transfer stations for bike-sharing commuters

In Fig.14, we aggregate the daily metro station usage for *Biking-transit* and *Transit-biking commuters*, with station size on the maps indicating the number of bike-sharing commuters. Notably, Fig.14 also contains *Biking-transit-biking commuters*, as they have the characteristics of both *Biking-transit* and *Transit-biking commuters*.

For *Biking-transit commuters* (Fig.14(a)), while the spatial distribution of metro stations resembles the residential hotspots shown in Fig.13(a), high bike-to-transit usage rates are concentrated in the suburban areas near the central city (e.g., Gushu, Minzhi, and Hongshan). This result aligns with previous studies (Guo et al., 2021; Zhang et al., 2024), revealing the primary residential distribution of the group using bike-sharing transfer services for across-district commuting. As for *Transit-biking commuters* (Fig.14(b)), most metro stations with high transit-to-bike usage rate are clustered in central areas near employment hubs (especially Nanshan and Futian districts). However, only a few metro stations exceed 900 daily transit-to-bike trips, likely due to the high accessibility within the central area and proximity of companies to metro exits (e.g., High-tech Park, Keyuan, etc.), which allows people to walk to work directly, thereby reducing the need for bike-sharing. Moreover, we observe that some metro stations (e.g., Bihaiwan, Gushu, Xili, etc.) exhibit both high numbers of both *Biking-transit and Transit-biking commuters*, which reflects a mixed use of living and working spaces in these areas. Based on these findings, we recommend that for stations with high commuting demand for bike-to-transit trips, bike-sharing operators should promptly redistribute bikes in the surrounding area to prevent their accumulation and encroachment on other road space. For stations with significant demand for transit-to-bike trips, improving bike reallocation efficiency is essential to avoid bicycle shortage during peak commuting hours. At stations with substantial demand of both bike-to-transit and transit-to-bike usage, data-driven riding analysis should be conducted, and precise scheduling strategies should be developed to accommodate the dynamic fluctuations in bike demand. For urban transportation planners, it is necessary to increase dedicated bike lanes, enhance traffic signage for better visibility, and provide ensure bicycle parking spaces around metro stations with high transfer demand. These measures would enhance residents' perception of the cycling environment, and help reduce uncivil behaviors, such as improper bike parking around metro stations.

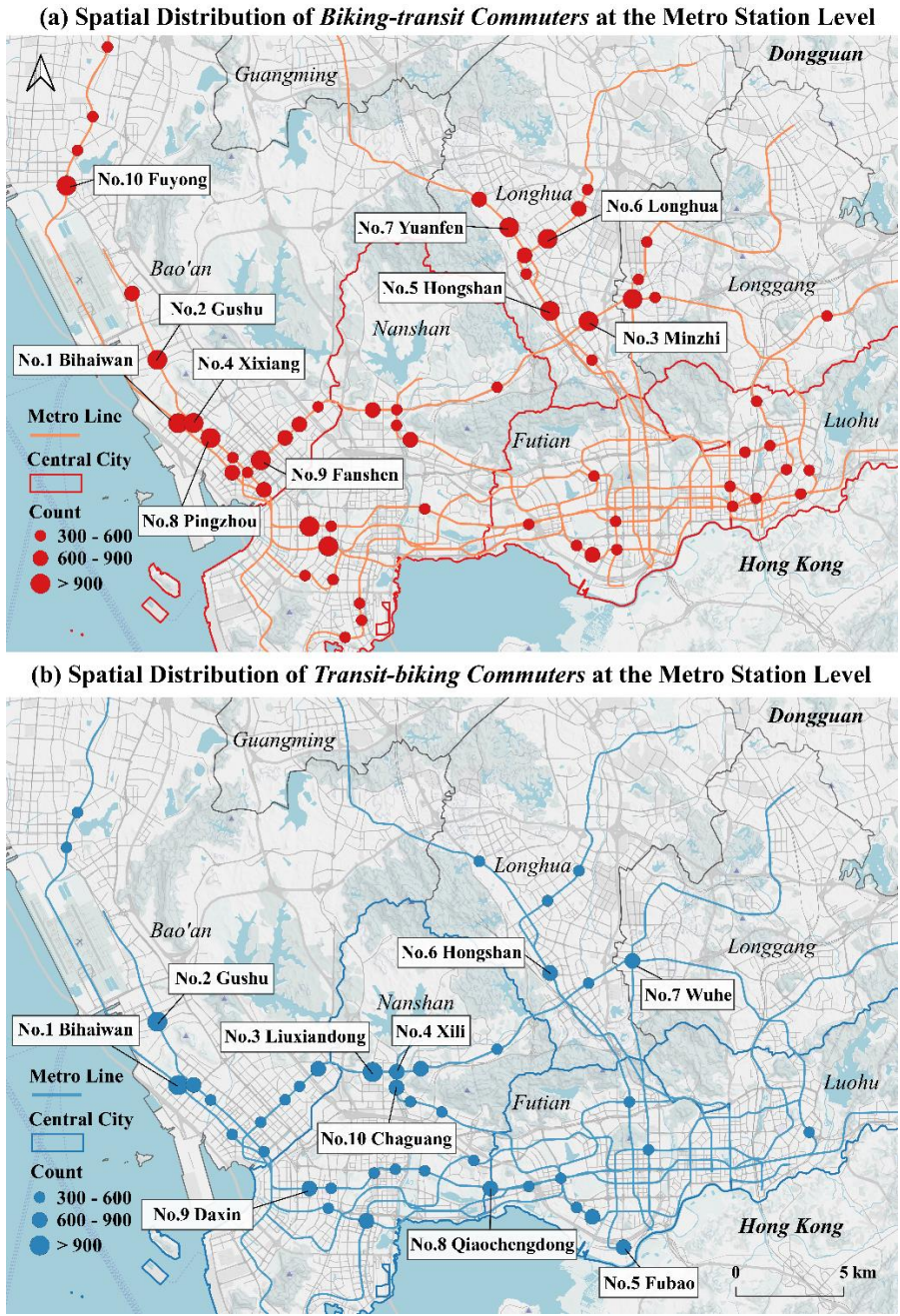

**Fig. 14** Spatial distribution of *Biking-with-transit commuters* at the metro station level

### 4.3.4 Spatial patterns of bike-sharing users' commuting chains

To gain further insights into the commuting mobility of bike-sharing users, we analyze the biking commuting chains for *Only-biking* and *Biking-transit-biking commuters* by linking their residences and workplaces to delineate their daily commuting flows, with home location as the origin and work location as the destination. Utilizing the spatial clustering method by X. Gao et al. (2020), we present the results of commuting flow clusters in Fig. 15 and 16.

For *Only-biking commuters* (Fig.15), we discover that the commuting flow clusters are generally have short distances, with an average length of 1.28 km. These flows consistently converged from residential hotspots to the nearest employment centers, in agreement with the observations in Fig. 10(b) and Fig.13. This result suggests that dockless bike-sharing play a significant role in short-distance commuting for both inner-city and suburban residents, further extending the findings of previous studies (S. Li et al., 2021; F. Gao et al., 2022). As for *Biking-transit-biking commuters* (Fig.16), the commuting flow clusters primarily extend from the suburbs to the central city, with an average length exceeding 15 km. Specifically, these users mostly reside in Bao'an and Longhua districts and daily use bike-sharing to transfer to the metro lines that connect the suburban and central areas (especially the Shenzhen Metro Lines 1, 4, 5, and 11). This finding corresponds to the actual situation in Shenzhen (e.g., many tech workers live near Pingzhou Station and work in the High-tech Park) and aligns with the analysis in Fig.14(a).

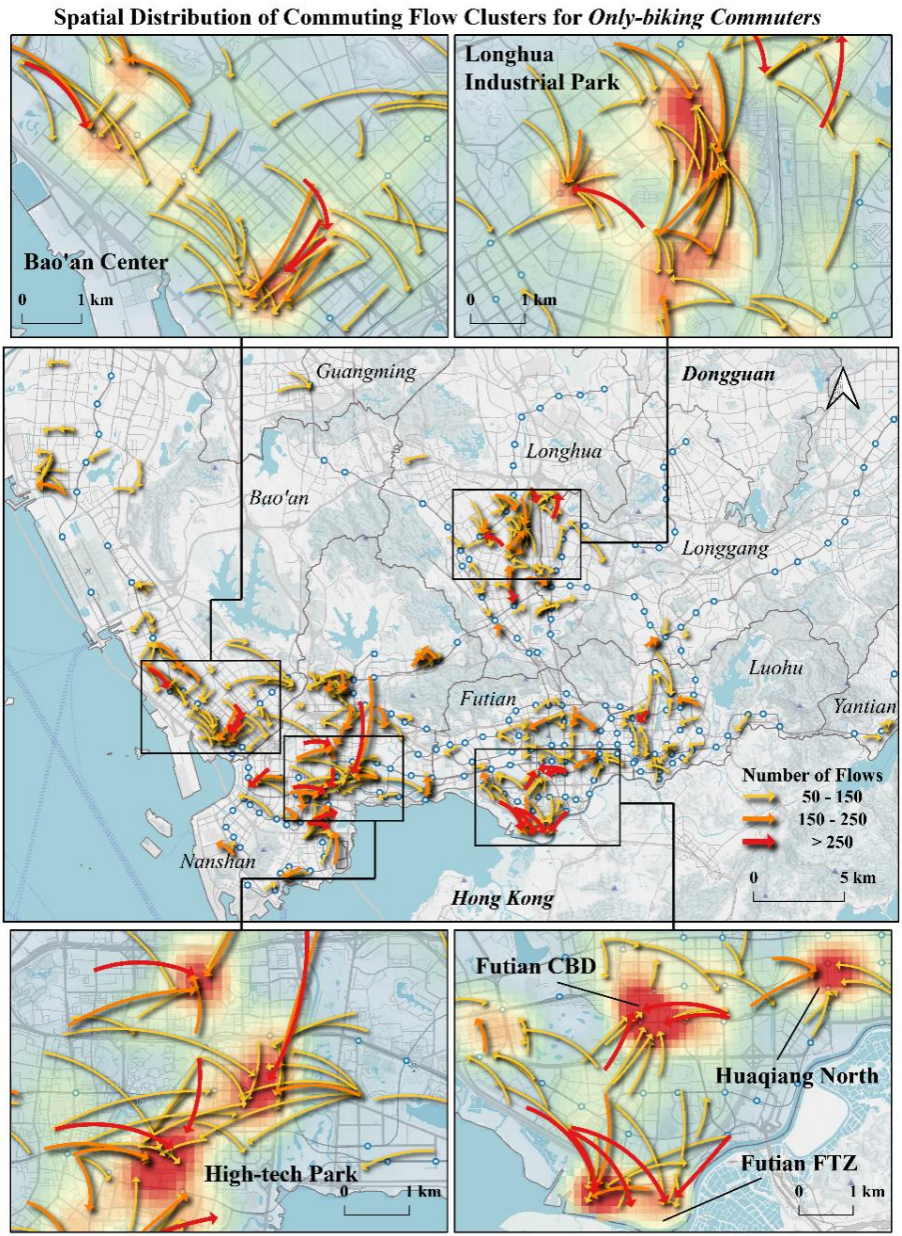

**Fig. 15** Spatial distribution of commuting flow clusters for *Only-biking commuters*

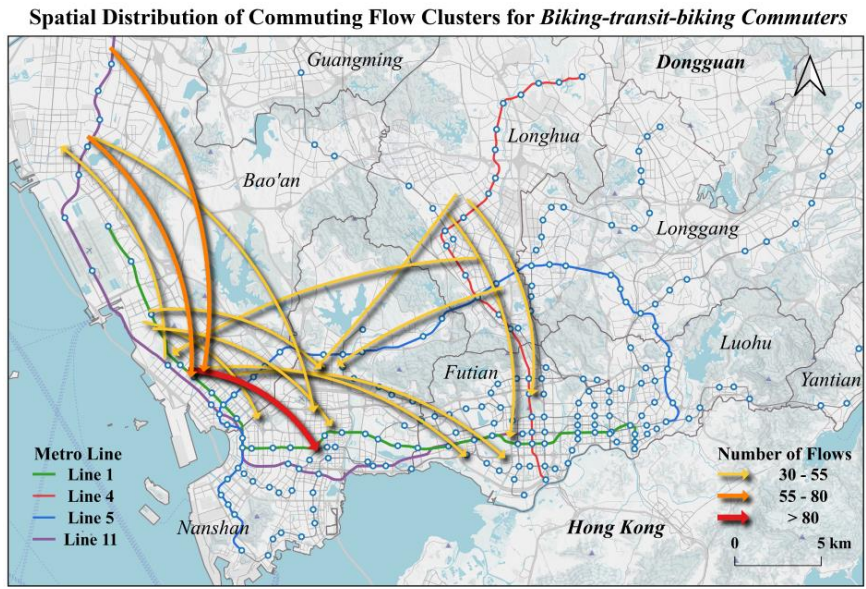

**Fig. 16** Spatial distribution of commuting flow clusters for *Biking-transit-biking commuters*

## 5 Conclusion

Mining daily trip chains of bike-sharing users at the individual level is crucial for the increasingly refined planning of active transportation, but it remains a complex task that has received limited attention in studies related to bike-sharing trip. To bridge this challenge, this paper presents a two-layer framework that integrates spatiotemporal flow clustering and rule-based decision trees, which is validated and applied to a dataset of over 200 million dockless bike-sharing records in Shenzhen. In Layer 1, to overcome the lack of geocoding in dockless bike-sharing dataset, we propose a flow clustering method with improved spatiotemporal constraints to identify users' daily trajectories from their disordered biking records. Its performance and applicability are confirmed through a comparative analysis with original clustering method. To the best of our knowledge, this is the first attempt to extract individual daily trajectories using spatiotemporal flow clustering method, which can be extended to relevant studies on other travel data with similar data features (e.g., taxi trip data). In Layer 2, considering the absence activity semantics in the individual trajectories extracted in Layer 1, we integrate round trip, working hours, and public transport transfer to construct three rule-based decision trees. These decision trees can identify the commuting behavior from each user's daily cycling trajectories to derive individual daily low-carbon commuting chains (including biking-with-transit). Such information can assist urban planners and bike-sharing operators to rapidly understand residents' daily cycling behaviors and demands. Moreover, by integrating multi-source data (e.g., street view images and housing prices), it can serve as a data foundation for refining individual cycling user profiles and conducting fine-scale research on bicycle behavior.

By applying the two-layer framework to the case study of Shenzhen, we have obtained some encouraging findings. First, the residential and workplace locations of bike-sharing commuters exhibit mixed distribution pattern, characterized by both widespread dispersion and local concentrations. The majority of bike-sharing commuters live in the urban villages and old communities (especially in central city), while a larger proportion of *Biking-with-transit commuters* reside in the suburban areas near the central city (e.g., the neighbourhood of Gushu and Hongshan metro stations). Second, some bike-sharing users show noticeable overtime patterns, with a higher proportion of *Only-biking commuters* compared to *Biking-with-transit commuters*, as the former incur lower commuting costs. Among the mainly job centers of the study area, Longhua Industrial Park, dominated by manufacturing, has the longest average working hours, exceeding 10 hours. Finally, we find that most active users utilize bike-sharing more frequently for commuting to work rather than returning home，, which is closely related to increased discretionary activities after work and the excessive bike-sharing demand around workplaces during commuting peak.

Based on the differences in daily commuting pattern observed among various categories of bike-sharing users, we emerge several corresponding policy implications. For *Only-biking commuters*, a significant portion of whom reside in suburbs and have longer working hours, we recommend that bike-sharing operators should focus on meeting bicycle demand during night-time hours (20:00-23:00) around the suburban industrial areas (e.g., Longhua Industrial Park). Meanwhile, we suggest that transportation planners could prioritize improvements to lighting infrastructure in these areas to enhance the safety of bicyclists commuting at night. Moreover, for high-demand metro stations with varying types of biking-transit connections, we propose that bike-sharing operators develop dynamic bike-sharing reallocation strategies based on the historical usage patterns at each station, to minimize mismatches between supply and demand. In parallel, urban planners should expand bike parking spaces around these high-demand stations and improve surrounding bicycle traffic signage to enhance residents' cycling perception and encourage more responsible bike-sharing practices.

However, there are still some limitations that warrant further improvement in future research. First, our framework limited to weekday commuting patterns of bike-sharing users, not accounting for weekend trips or non-commuting activities (such as exercising and leisure). Subsequent studies can leverage location service data (e.g., points of interest) to explore the cycling activity in these contexts, or integrate other geotagged big data (e.g., smart card data) to develop more detailed low-carbon travel chain models (Zhang et al., 2024). Second, it is necessary to validate mobility patterns with travel survey data. Unfortunately, due to the difficulty in obtaining relevant data covering the cycling population in the study area, this is

48/58

not possible in our research. Moreover, our study only selected Shenzhen as the case study area. Future research could include more cities to compare the differences in cycling mobility patterns and explore their relationship with urban features (e.g., built environment). Lastly, noted that in many megacities in China, private bicycles (especially electric-bike) still play a role in transportation. Investigating whether their mobility patterns are similar to those of bike-sharing users would be valuable, as it pertains to maximizing the benefits of building cycling-friendly environments.